\documentclass[journal=nalefd, manuscript=letter, email=true, layout=twocolumn]{achemso}
\usepackage{chemformula} % Formula subscripts using \ch{}
\usepackage[T1]{fontenc} % Use modern font encodings
\usepackage{caption}
\usepackage{subcaption}
\usepackage{siunitx}
\DeclareSIUnit\angstrom{\protect \text {Å}}
\usepackage{physics}
\usepackage{amsmath, amssymb}
\usepackage[version=4]{mhchem}
\setkeys{acs}{doi = true}
\usepackage[nobiblatex]{xurl}
\usepackage{xcolor}

\mciteErrorOnUnknownfalse

\makeatletter
\renewcommand*{\acs@author@fnsymbol@symbol}[1]{% Use numbers instead of symbols, * is for email
    \ifcase #1 *\or
    1\or
    2\or
    3\or
    4\or
    5\or
    6\or
    7\or
    8\or
    9\or
    10
    \fi
}
        
\renewcommand*\acs@contact@details{% addd * before  E-mail
    {\sffamily *\,E-mail: \acs@email@list }%
    \acs@number@list
}           

\patchcmd{\acs@address@list@auxii}% superscript for numbers in affiliations
{\acs@author@fnsymbol{\acs@affil@marker@cnt}}
{\textsuperscript{\acs@author@fnsymbol{\acs@affil@marker@cnt}}}
{}{}

\patchcmd{\acs@address@list@auxii}% superscript for numbers in affiliations
{{\acs@author@fnsymbol{\acs@affil@marker@cnt}\@nameuse{@altaffil@\@roman\@tempcnta}\par}}
{{\textsuperscript{\acs@author@fnsymbol{\acs@affil@marker@cnt}}\@nameuse{@altaffil@\@roman\@tempcnta}\par}}
{}{}        

\makeatletter % smaller figure captions
\long\def\@makecaption#1#2{%
  {
   \footnotesize % or whatever your desired size is
   #1: #2
  }}
\makeatother

\makeatother    
\let\oldmaketitle\maketitle
\let\maketitle\relax

\author{\normalsize Lorenzo Graziotto}
\email{lgraziotto@phys.ethz.ch}
\affiliation{\footnotesize Department of Physics, Sapienza University of Rome, Piazzale Aldo Moro 5, 00185 Rome, Italy}%#
\altaffiliation{Current Address: Institute of Quantum Electronics, ETH Zürich, Auguste-Piccard-Hof 1, 8093 Zürich, Switzerland}
\author{Francesco Macheda}
\affiliation{\footnotesize Department of Physics, Sapienza University of Rome, Piazzale Aldo Moro 5, 00185 Rome, Italy}
\alsoaffiliation{Istituto Italiano di Tecnologia, Graphene Labs, Via Morego 30, 16163 Genoa, Italy}
\author{Tommaso Venanzi}
\affiliation{\footnotesize Department of Physics, Sapienza University of Rome, Piazzale Aldo Moro 5, 00185 Rome, Italy}%#
\author{Guglielmo Marchese}
\affiliation{\footnotesize Department of Physics, Sapienza University of Rome, Piazzale Aldo Moro 5, 00185 Rome, Italy}%#
\author{Simone Sotgiu}
\affiliation{\footnotesize Department of Physics, Sapienza University of Rome, Piazzale Aldo Moro 5, 00185 Rome, Italy}%#
\author{Taoufiq Ouaj}
\affiliation{JARA-FIT and 2nd Institute of Physics, RWTH Aachen University, 52074 Aachen, Germany}
\author{Elena Stellino}
\affiliation{Department of Physics and Geology, University of Perugia, Via Alessandro Pascoli, 06123 Perugia, Italy}
\author{Claudia Fasolato}
\affiliation{Institute for Complex System, National Research Council (ISC-CNR), 00185 Rome, Italy}
\author{Paolo Postorino}
\affiliation{\footnotesize Department of Physics, Sapienza University of Rome, Piazzale Aldo Moro 5, 00185 Rome, Italy}
\author{Marvin Metzelaars}
\affiliation{Institute of Inorganic Chemistry, RWTH Aachen University, 52074 Aachen, Germany}
\author{Paul K{\"o}gerler}
\affiliation{Institute of Inorganic Chemistry, RWTH Aachen University, 52074 Aachen, Germany}
\author{Bernd Beschoten}
\affiliation{JARA-FIT and 2nd Institute of Physics, RWTH Aachen University, 52074 Aachen, Germany}
\author{Matteo Calandra}
\affiliation{Department of Physics, University of Trento, Via Sommarive 14, 38123 Povo, Italy}
\author{Michele Ortolani}
\affiliation{\footnotesize Department of Physics, Sapienza University of Rome, Piazzale Aldo Moro 5, 00185 Rome, Italy}
\author{Christoph Stampfer}
\affiliation{JARA-FIT and 2nd Institute of Physics, RWTH Aachen University, 52074 Aachen, Germany}
\author{Francesco Mauri}
\affiliation{\footnotesize Department of Physics, Sapienza University of Rome, Piazzale Aldo Moro 5, 00185 Rome, Italy}
\alsoaffiliation{Istituto Italiano di Tecnologia, Graphene Labs, Via Morego 30, 16163 Genoa, Italy}%
\author{Leonetta Baldassarre}
\affiliation{\footnotesize Department of Physics, Sapienza University of Rome, Piazzale Aldo Moro 5, 00185 Rome, Italy}%#

\title[]{\large Infrared resonance Raman of bilayer graphene: signatures of massive fermions and band structure on the 2D peak}
\begin{document}

%\begin{tocentry}

%Some journals require a graphical entry for the Table of Contents.
%This should be laid out ``print ready'' so that the sizing of the

%Inside the \texttt{tocentry} environment, the font used is Helvetica
%8\,pt, as required by \emph{Journal of the American Chemical

%The surrounding frame is 9\,cm by 3.5\,cm, which is the maximum
%permitted for  \emph{Journal of the American Chemical Society}
%graphical table of content entries. The box will not resize if the
%content is too big: instead it will overflow the edge of the box.

%This box and the associated title will always be printed on a
%separate page at the end of the document.

%\end{tocentry}

\twocolumn[
\begin{@twocolumnfalse}
\oldmaketitle
\begin{abstract}
 Few-layer graphene possesses low-energy carriers which behave as massive fermions, exhibiting intriguing properties in both transport and light scattering experiments. Lowering the excitation energy of resonance Raman spectroscopy down to \SI{1.17}{eV} we target these massive quasiparticles in the split bands close to the \textbf{K} point. The low excitation energy weakens some of the Raman processes which are resonant in the visible, and induces a clearer frequency-separation of the sub-structures of the resonance 2D peak in bi- and trilayer samples. We follow the excitation-energy dependence of the intensity of each sub-structure and, comparing experimental measurements on bilayer graphene with \emph{ab initio} theoretical calculations, we trace back such modifications on the joint effects of probing the electronic dispersion close to the band splitting and enhancement of electron-phonon matrix elements.
 %in the case of bilayer graphene we unveil an enhanced coupling between the massive fermions and the lattice vibrations at the \textbf{K} point, in analogy to what found for the massless fermions of monolayer graphene, and also suggesting that what governs the enhancement is the vicinity of the electron-hole pair momentum to \textbf{K} rather than how small the electron-hole pair energy is.
\end{abstract}
 \vspace{0.5cm}
\end{@twocolumnfalse}
]

Raman scattering can be used as a powerful experimental tool to explore the electron–phonon interaction in two-dimensional (2D) materials~\cite{loudon1965theory}. The strength of the electron-phonon coupling (EPC), which is determined by the interplay between atomic displacements and the electronic band structure of the material, can be strongly modified in 2D materials where, due to the reduced dimensions, long-range Coulomb interactions are less efficiently screened~\cite{PhysRevB.66.235415, PhysRevB.80.224301, attaccalite2010doped, Sohier2017, macheda2023elecphon}. By resorting to resonance Raman scattering, i.e.\ Raman processes involving an electronic transition between two real states of the system under exam, one can obtain high Raman signal also for several modes with non-zero $\vb{q}$ wavevectors within the first Brillouin zone (BZ)~\cite{cardona1982light}.
%In several 2D systems it has been shown that second-order double resonant bands can arise~\cite{sotgiu2022mose, carvalho2017intervalley}, stemming from inter-valley or intra-valley scattering of the electrons and holes emitting phonons either at zone boundary ($\vb{q} \sim \mathbf{K}$) or zone center ($\vb{q} \sim \boldsymbol{\Gamma}$). 
For example, the Raman spectrum of graphene is composed by a first-order mode at $\vb{q} = 0$ (the G peak) and several double-resonance modes (called D, D+D'', 2D, 2D'), which have wavevector $\vb{q} \neq 0$~\cite{ferrari2006raman, malard2009raman}. By changing the incoming laser energy ($\epsilon_L = \hbar\omega_L = hc/\lambda_L$), one probes different regions of both the electron and phonon dispersion, and any variation in the electronic properties (due to number of layers, doping, defects, or strain) is reflected into a modification of the position, the width and the intensity of the Raman peaks~\cite{venezuela2011theory, graf2007spatially, ferreira2010evolution, neumann2015raman}. 

In Refs.~\cite{basko2008interplay, lazzeri2008impact} it was theoretically predicted that in graphene the coupling between carriers in the vicinity of the Dirac point and zone-boundary phonons gets strongly enhanced by Coulomb interactions. Given this theoretical prediction, valid in general for any 2D material displaying Dirac cones, it is of great interest to perform resonance Raman experiments with laser emitting in the infrared, in order to excite electron-hole pairs as close as possible to the Dirac point. Indeed, recently~\cite{venanzi2023probing}, experiments performed on monolayer graphene with  laser excitation energies down to $\epsilon_L=\SI{1.17}{eV}$ have shown a large enhancement of the ratio between the 2D and the 2D' resonance Raman peak intensities, which was explained in terms of a momentum-dependent EPC that is enhanced for carriers interacting with phonons at $\mathbf{K}$ with respect to those at $\mathbf{\Gamma}$. In a similar fashion, an enhanced EPC could occur also for the massive fermions of bilayer graphene, where the resonance condition near the bottom of the split bands can be matched~\cite{malard2007probing} with infrared light. 
%. Such intensity ratio is proportional to the fourth power of the electron phonon coupling (EPC) at $\vb{q} \sim \mathbf{K}$ divided by the fourth power of EPC at $\vb{q} \sim \boldsymbol{\Gamma}$. %In bilayer graphene, where the resonance condition near the bottom of the split bands can be matched~\cite{malard2007probing} with infrared light, one could therefore in principle study the if there is any modification of the resonant scattering pathways of electron-hole pairs where the wavevector of the exchanged phonon $\vb{q}$ almost equals $\mathbf{K}$, and compare this to quat has been found in monolayer graphene. %In bilayer graphene, the 2D peak for visible excitation energies has been described and interpreted with four main scattering processes~\cite{herziger2014bilayer}. 

In this work we show how the line-shape of the 2D peak for bilayer and trilayer graphene is modified with near-infrared (NIR) excitation energy, compared to the line-shape for visible (VIS) excitation energy. We find that for bilayer graphene such line-shape modification corresponds to  a rearrangement of the intensity among the different sub-peaks that compose the resonance peak. 
We compare the resonance Raman spectra on hBN-encapsulated bilayer graphene to \emph{ab initio} calculations, and prove that we can spectrally resolve and identify the different resonant scattering pathways within the 2D peak, attempting to understand the origin of the spectral weight shift within the different scattering processes.
\\
%We find that for bilayer graphene the modification of the lineshape with decreasing laser excitation energy corresponds to the a modification of the relative intensities among the different sub-features and from the analysis of the . Our calculations can however not fully reproduce the peak relative intensity for $\hbar\omega_L$ =1.17 eV and we discuss by state-of-the art calculations due to (i) an incorrect estimation of the gap width between the split bands, (ii) a momentum-dependent electron-phonon coupling (underestimated in DFT calculations), and (iii) an enhancement of the electron-light matrix element. \\
 
%\section{Results and discussion}

%\textbf{Evolution of the 2D peak line-shape with excitation energy and number of layers---} 
We have performed resonance Raman spectra measurements on hBN-encapsulated few-layer graphene for a number of laser excitation energies (wavelengths):\\

\noindent
\vspace{0.5cm}
\resizebox{\linewidth}{!}{%
\begin{tabular}{c!{\vrule width 0.1em}c|c|c|c|c|c}
    
    $\epsilon_L$ [\SI{}{eV}] &  1.17 & 1.58 & 1.96 & 2.33 & 2.54 & 3.06 \\[0.2\normalbaselineskip]
    \hline \rule{0pt}{1.\normalbaselineskip}%
    $\lambda_L$ [\SI{}{nm}] & 1064 & 785 & 633 & 532 & 488 & 405 
\end{tabular}}

\noindent
For laser excitation energy $\epsilon_L \geq \SI{1.58}{eV}$ we have relied on dispersive Raman setups (\textit{Horiba LabRAM HR}, and \textit{WiTec} for \SI{1.58}{eV}), while $\epsilon_L = \SI{1.17}{eV}$ spectra have been collected with a Michelson interferometer coupled via fibers to an optical microscope (\textit{Bruker MultiRAM}). The spectra have been corrected for the CCD response function or for that of the single point germanium detector. The spectral calibration for \SI{1.17}{eV} data is ensured by the interferometric detection scheme, while the other data have been spectrally calibrated with Neon light spectra.

\begin{figure}[hbt!]
  \includegraphics[width=\linewidth]{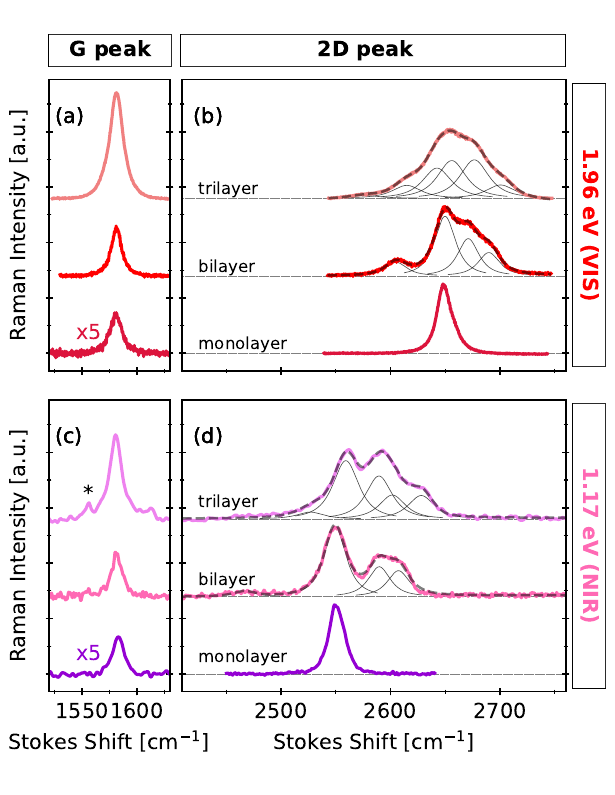}
  \caption{G \textbf{(a-c)} and 2D \textbf{(b-d)} Raman peaks measured with visible (VIS) $\epsilon_L = \SI{1.96}{eV}$ \textbf{(a-b)} and near-infrared (NIR) $\epsilon_L = \SI{1.17}{eV}$ \textbf{(c-d)} for monolayer, bilayer and trilayer graphene. Each curve has been normalized to the respective height of the 2D peak, and the intensity of the G peak of the monolayer has been multiplied by five for visual clarity. The star near trilayer's G peak at \SI{1.17}{eV} indicates a spurious Raman peak due to oxygen molecules. Notice how the shape of the G peak does not vary across the different number of layers and excitation energies, while the 2D peak is strongly modified, developing, as the number of layers is increased, features at both higher and lower Stokes shift. These features are fitted via the sum of 4 (bilayer)~\cite{herziger2014bilayer} or 6 (trilayer)~\cite{cong2011raman} Baskovian functions (see Supplemental Material). Lowering the excitation energy to $\epsilon_L = \SI{1.17}{eV}$, these features become better separated in frequency and some of them are strongly suppressed, in particular in the case of bilayer graphene, where just three Baskovian functions are used.}
  \label{fig:G_2D_monobitri}
\end{figure}

\begin{figure}[!htb]
  \includegraphics[width=\linewidth]{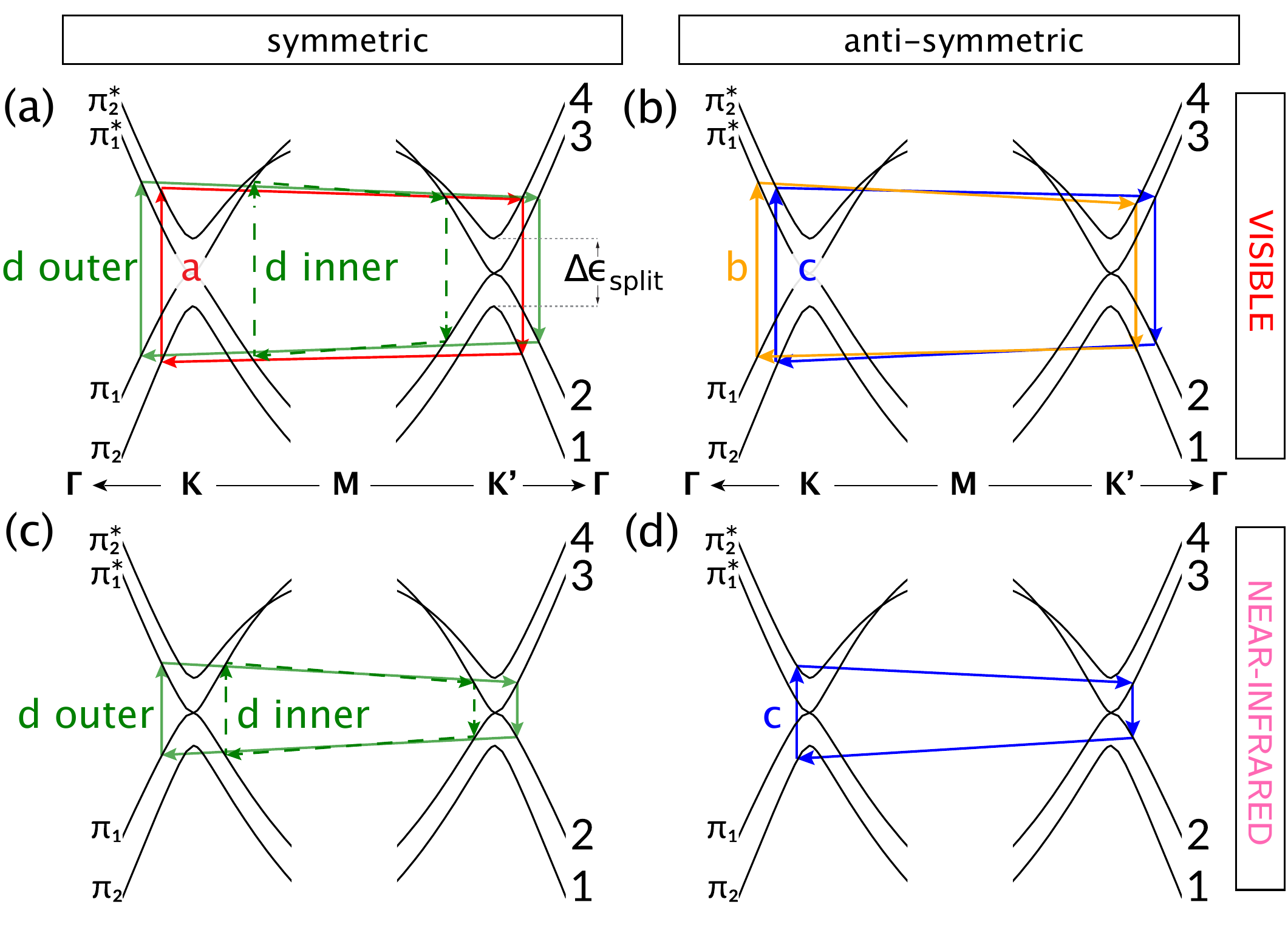}
  \caption{\textbf{(a-d)} Diagrams of the most significant scattering processes that build up the 2D peak in bilayer graphene for visible \textbf{(a-b)} and near-infrared \textbf{(c-d)} excitation energy, involving the electronic bands along the high-symmetry $\boldsymbol{\Gamma}-\mathbf{K}-\mathbf{M}$ direction. The labelling 1-2-3-4 of the bands follows their energy ordering at the \textbf{K} point, while the labelling of the processes separates them in symmetric (processes a and d), where the electron/hole is scattered within the same band, and anti-symmetric (processes c and b), where instead the electron/hole changes band upon scattering with the TO phonon. For illustrative purposes we show only the outer contribution, except for the D process where we further distinguish between inner and outer contributions.}
  \label{fig:schemeProcesses}
\end{figure}

In Figure~\ref{fig:G_2D_monobitri} we report Raman measurements for $\epsilon_L = 1.96$ and \SI{1.17}{eV} on monolayer, bilayer and trilayer graphene. As expected the 2D peak line-shape of bi- and trilayer graphene is built up by several different sub-peaks, which we deconvolve with the sum of four (bilayer) or six (trilayer) \emph{Baskovian}~\cite{basko2008theory} functions $f_B(\omega)$ (see Supplemental Material (SM) for definition). We observe in particular that the 2D peak of bilayer graphene is dramatically modified by lowering the laser excitation energy, as it displays two main features well separated in frequency for $\epsilon_L = \SI{1.17}{eV}$, which we fit via the sum of three Baskovian functions. The overall 2D peak line-shape of trilayer graphene for $\epsilon_L = \SI{1.17}{eV}$ is also modified with respect to the one for $\epsilon_L = \SI{1.96}{eV}$ with a larger distance in energy between the center frequencies of its sub-peaks, allowing for a better deconvolution of its line-shape, showing four dominant sub-peaks instead of six. This spectral evolution is explained in terms of the electronic bands dispersion proper of the different types of few-layer graphene~\cite{gruneis2008tight, mak2010evolution, coletti2013revealing, popov2015two}. In Bernal-stacked bilayer graphene the 2p$_z$ orbitals of the four carbon atoms in the unit cell hybridize to give rise to two pairs of valence ($\pi_1$, $\pi_2$) and conduction ($\pi_1^*$, $\pi_2^*$) bands~\cite{mccann2013electronic} (depicted in each panel of Figure~\ref{fig:schemeProcesses}). In Bernal-stacked trilayer graphene one finds instead three pairs of valence ($\pi_1$, $\pi_2$, $\pi_3$) and conduction bands ($\pi_1^*$, $\pi_2^*$, $\pi_3^*$)~\cite{torche2017first} (see SM). Although bilayer or trilayer graphene may be viewed just as two or three coupled monolayers, and indeed in the intrinsic case they possess zero band gap, the low-energy electronic dispersion bears striking differences with respect to the monolayer: the $\pi_1$ and $\pi_1^*$ bands, which are degenerate at the \textbf{K} and \textbf{K'} points (where in the undoped case the Fermi level lies), display a quadratic behaviour with massive quasiparticles, at variance with the massless fermions in monolayer's Dirac cones. Moreover, the second pair of bands ($\pi_2$, $\pi_2^*$ in bilayer, $\pi_3$, $\pi_3^*$ in trilayer), which still shows quadratic dispersion, is split at \textbf{K} by an energy difference which is estimated to be about $\Delta\epsilon_\mathrm{split} = \SIrange{0.7}{0.8}{eV}$ in bilayer~\cite{kuzmenko2009infrared, PhysRevB.78.235408} and \SI{1.2}{eV} in trilayer~\cite{cong2011raman}. In Bernal-stacked trilayer graphene the third pair of bands ($\pi_2$, $\pi_2^*$) shows the same Dirac cone dispersion of the monolayer~\cite{bao2011stacking}. These tangled band structure modifications are mapped into the sub-peaks that build up the 2D peak line-shape. The latter arise due to the different resonant scattering pathways between neighbouring cones, as pointed out in Ref.~\cite{herziger2014bilayer, torche2017first}, and are evident both in the case of bilayer and trilayer graphene.

\begin{figure*}[!hbt]
\hspace{-0.55 cm}
  \includegraphics[width=1.\linewidth]{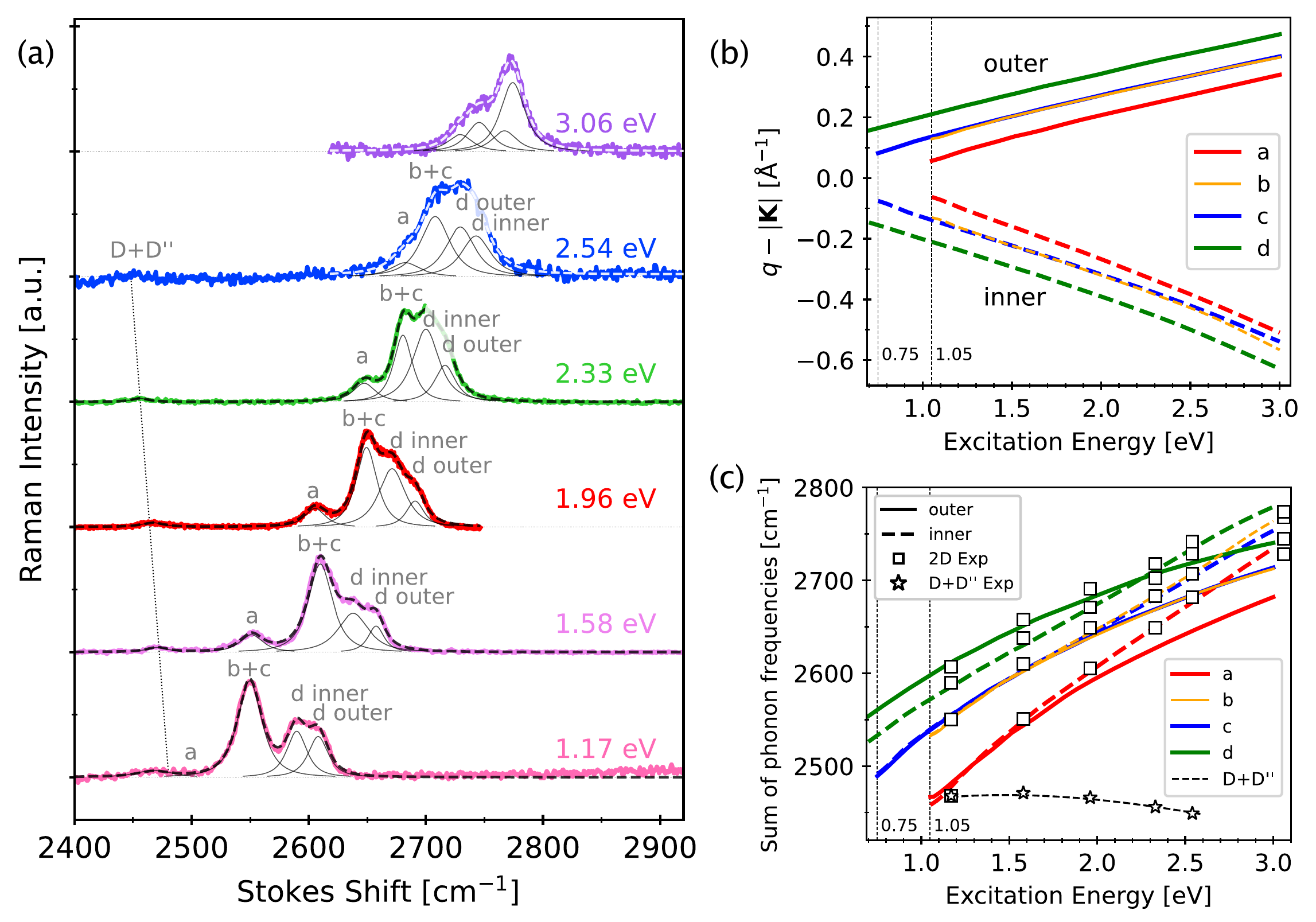}
  \caption{\textbf{(a)} Experimental evolution of the 2D peak line-shape in bilayer graphene for different laser excitation energies. The D+D'' resonance peak is also shown at lower Stokes shift, and its dispersion can be followed with the guide-to-the-eye dashed line. Notice in particular the dispersion of the 2D peak as a whole (moving in the opposite direction with respect to the D+D'' peak), and the change in spectral weight between its components. Each spectrum has been normalized to its maximum, and the 2D sub-features have been fitted via the sum of four $f_B$ functions. For $\epsilon_L < \SI{3.0}{eV} $ the processes referring to the experimental sub-features are labelled: notice that the b and c processes constructively interfere since they involve the same phonons, and for the d process one can distinguish between the contributions of inner and outer phonons (which swap places for $\epsilon_L > \SI{2.4}{eV}$) as discussed in the main text. For higher excitation energies (i.e.\ \SI{3.06}{eV}) it becomes harder to attribute a label to each $f_B$ since processes other than the a, b, c, d described in Fig.~\ref{fig:schemeProcesses} become relevant, and constructively interfere with each other to build up the contribution at higher Stokes shift. Finally, at $\epsilon_L = \SI{1.17}{eV}$ the a sub-peak of the 2D peak and the D+D'' peak share similar Stokes shift, and are difficultly separated experimentally. \textbf{(b)}~Resonant phonon wavevector $\vb{q}$ along $\boldsymbol{\Gamma}-\mathbf{K}-\mathbf{M}$ as a function of $\epsilon_L$, for the different a, b, c, d processes and inner/outer contributions, as obtained via \emph{ab initio} calculations. Notice that for $\epsilon_L < \SI{1.05}{eV}$ the processes a and b become virtual, since the scattered electron-hole pair cannot recombine resonantly on the split 1-4 bands. Moreover, below the band-gap $\Delta\epsilon_\mathrm{split} = \SI{0.75}{eV}$ also process c is suppressed. Notice also that inner and outer $\vb{q}$ process wavevectors become almost symmetric with respect to \textbf{K} as $\epsilon_L$ is reduced, since the dispersions approach an isotropic shape. \textbf{(c)}~Total frequency $\Omega_\mathrm{2D}$ of the scattered phonon pair with wavevector $\vb{q}$ specified in panel \textbf{b}, as obtained via \emph{ab initio} calculations for both inner and outer processes (dashed and solid line, respectively), and offset in order to match the experimental data. The square markers represent the central frequencies of the $f_B$ functions with which the experimental data of panel \textbf{a} have been fit. The star markers represent instead the frequency of the D+D'' peak. Notice that for $\epsilon_L < \SI{2.0}{eV}$ process d has the inner and outer contributions considerably split in frequency, and that the agreement with the theoretical curves becomes less accurate as $\epsilon_L$ is increased, since the simplified inner/outer separation along $\boldsymbol{\Gamma}-\mathbf{K}-\mathbf{M}$ is not sufficient to describe the shape of the peak.}
  \label{fig:2Dpeak_exp}
\end{figure*}

We now focus on the bilayer graphene case, where at $\epsilon_L = \SI{1.17}{eV}$ the separation between sub-peaks is more evident, and we show that the three Baskovian sub-peaks correspond mainly to two resonant scattering pathways. In order to lighten the notation, in the following we will refer to the four low-energy electronic bands of bilayer graphene by enumerating them in ascending order (with respect to the energy at the \textbf{K} point), as labelled on the right side of Fig.~\ref{fig:schemeProcesses}a. As already understood in Refs.~\cite{ferrari2013raman, herziger2014bilayer}, the incident laser light of frequency $\omega_L$ couples predominantly the two pairs of corresponding bands, i.e.\ 1 and 4, or 2 and 3, generating electron-hole pairs either in bands 1-4 or in bands 2-3, with wavevectors close to the \textbf{K} point. Subsequently, both the electron and the hole are inelastically scattered by two quasi-degenerate TO phonons~\cite{herziger2014bilayer} (which have opposite wavevectors $\vb{q} \sim \mathbf{K}$ to satisfy crystal momentum conservation, and have total energy $\hbar \Omega_\mathrm{2D} \sim \SI{0.3}{eV}$ for $\epsilon_L = \SI{1.17}{eV}$, which depends on the excitation energy via the phonon dispersion, see Fig.~\ref{fig:2Dpeak_exp}c) to the neighbourhood of the \textbf{K'} point, either on the same conduction/valence band which they started in (symmetric processes, which are labelled a and d in Fig.~\ref{fig:schemeProcesses}a), or in the other one (anti-symmetric processes, labelled c and b in Fig.~\ref{fig:schemeProcesses}b). Finally, the electron and the hole recombine and a photon with frequency $\omega_L - \Omega_\mathrm{2D}$ is emitted. The notable aspect of double-resonance Raman processes is that all the intermediate states of the process are real states of the system, hence the electron-hole pair generation (recombination) must match the energy of the incoming (scattered) photon. This is indeed the case for the four processes which mainly contribute to the 2D peak in the case of visible excitation energies (as depicted in Fig.~\ref{fig:schemeProcesses}a and~\ref{fig:schemeProcesses}b). On the other hand, by lowering the excitation energy more to the infrared the electron-hole pair upon scattering on the TO phonon pair can match the frequency $\omega_L - \Omega_\mathrm{2D}$ of the scattered photon only if it belongs to the pair of \textbf{K}-degenerate bands 2-3 or if
it has an energy greater than $\Delta\epsilon_\mathrm{split}$. Thus, while processes c and d are resonantly allowed, a and b are sensitive to the value of $\Delta\epsilon_\mathrm{split}$ (see SM) and display a strong suppression compared to the case of visible excitation energy (Fig.~\ref{fig:schemeProcesses}c and~\ref{fig:schemeProcesses}d). With this understanding, we use \SI{1.17}{eV} excitation energy with the aim of suppressing the contribution of a and b processes to the 2D peak, and then clearly separate in frequency the contributions from c and d (see Fig.~\ref{fig:G_2D_monobitri}d for bilayer and the discussion below). We point out that, while it is customary to resolve the experimental peaks by fitting them with the sum of more than one $f_B$ (as done in Fig.~\ref{fig:G_2D_monobitri}), these peaks do not immediately refer to the a, b, c, d processes described above. Indeed, not only the resonance condition selects phonons with some energy dispersion (determined by the mutual interplay of the electronic and phononic dispersion details~\cite{herziger2014bilayer, graziotto2023raman}), but also different processes involving the same phonon can interfere, so that the total intensity will be different from the mere sum of their squared amplitudes. For each of the a, b, c, d processes, a simplified treatment~\cite{berciaud2013intrinsic} addresses the first issue by further distinguishing between \emph{inner} and \emph{outer} processes (see Fig.~\ref{fig:2Dpeak_exp}b). Inner (outer) refers to the processes in which the exchanged phonon wavevector $\vb{q}$ lies along the $\boldsymbol{\Gamma}-\mathbf{K}-\mathbf{M}$ direction and it is smaller (greater) than \textbf{K}. This definition is extended to $\vb{q}$ belonging to the whole BZ in Ref.~\cite{herziger2014bilayer} (see SM). In the following we only address the experimental sub-features as the complex sum of the a, b, c, d processes, irrespective of the inner and outer distinction, and we compare to the full \emph{ab initio} calculations.~\cite{herziger2014bilayer, graziotto2023raman}. 

It is interesting to investigate experimentally and theoretically the relative intensity of the c and d Raman processes as a function of the laser energy, since for the c process $\epsilon_L=1.17$ eV generates electron-hole pairs in the split bands 2-3 close to the \textbf{K} point, and therefore could give us optical access to the EPC near the zone-boundary. In Figure~\ref{fig:2Dpeak_exp}a we display the experimental bilayer graphene 2D peak line-shape for several excitation energies between \SI{3.06}{} and \SI{1.17}{eV}. Below \SI{2500}{cm^{-1}} one can also identify the double-resonance D+D'' peak, which involves different phonons with respect to the ones of the 2D peak~\cite{PhysRevB.87.075402}, and blue-shifts with increasing $\epsilon_L$. Lowering the excitation energy, along with the overall red-shift of the 2D peak as a whole, which stems from the TO phonon dispersion, the most notable change is a shift of spectral weight within the peak. At $\epsilon_L=\SI{3.06}{eV}$ the most prominent sub-peak is the one at higher Stokes shift, associated to process d, while at \SI{1.17}{eV} the most relevant contribution comes from the sub-peak associated to processes b+c. For $\epsilon_L = \SI{1.96}{}$ and \SI{1.58}{eV} the sub-peak ascribed to the b+c processes becomes progressively more intense with respect to that ascribed to the d process, while the intensity related to the a process appears to remain fairly constant. At $\epsilon_L = \SI{1.17}{eV}$ the b+c sub-peak is more intense and well separated from the d sub-peak and it is very hard to identify the presence of a clear feature related to the a process alone, as it overlaps spectrally with the D+D'' peak. Remarkably, we find that for $\epsilon_L = \SI{1.17}{eV}$ the relative intensity of the b+c and d sub-peaks is dependent on the application of back-gate voltage or changing the substrate (see Fig.~S3 in the SM), suggesting that doping and modifications of $\Delta\epsilon_\mathrm{split}$ impact the 2D line-shape. One may suppose the intensity of the a and b processes to scale in a similar fashion when $\omega_L -\Omega_\mathrm{2D}$ becomes close to or smaller than $\Delta\epsilon_\mathrm{split}$. This in turn would suggest that the shift of spectral weight between the b+c and the d sub-peaks is driven by an enhancement of the c sub-peak. One possibility is thus that the enhancement of the c process could arise because the electron-hole pair is excited close to the band edge, due to the increased density of states. A small modification of the band structure could thus impact the intensity of the c process, much more than what happens for the d process. 

\begin{figure*}[hbt!]
  \includegraphics[width=\linewidth]{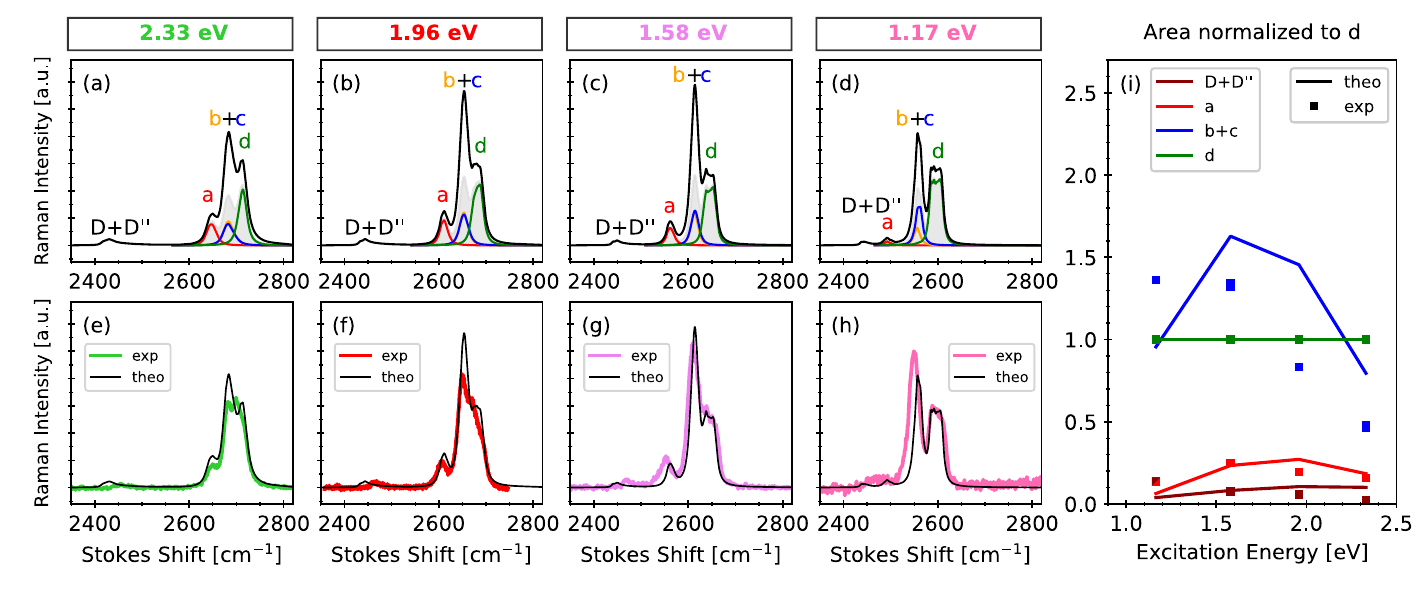}
  \caption{Comparison of the experimental bilayer 2D peak Raman spectra measured at $\epsilon_L =$ \SI{2.33}{}, \SI{1.96}{}, \SI{1.58}{}, and \SI{1.17}{eV} with the results of \emph{ab initio} calculations, where we added an offset to the Stokes shift in order to match the position in frequency. Each spectrum is normalized to the integrated area of its d sub-peak. \textbf{{(a-d)}} Theoretical spectra (black curve), with the peaks related to the processes described in Fig.~\ref{fig:schemeProcesses}a-d (which follow the same labelling and color code): the importance of the quantum interference between the processes is highlighted by the fact that the total intensity (the square of the sum of the amplitudes) is greater than the mere sum of the processes' intensities (grey shaded area). \textbf{(e-g)} Comparison between the theoretical spectrum (solid black) and the experimental 2D peak spectrum (green, red, violet, and pink curves, for $\epsilon_L =$ \SI{2.33}{}, \SI{1.96}{}, \SI{1.58}{}, and \SI{1.17}{eV}, respectively). The theoretical calculations are performed with $\Delta\epsilon_\mathrm{split} = \SI{0.75}{eV}$. Observe how the shape is reproduced fairly good in the case of $\epsilon_L = \SI{2.33}{}$, \SI{1.96}{} (as first shown in Ref.~\cite{herziger2014bilayer}), and \SI{1.58}{eV}, while for \SI{1.17}{eV} the calculation underestimates the intensity of the b+c processes. \textbf{(i)} Integrated area of the different processes shown in panels a-h normalized to process d, in experiment (squares) and \emph{ab initio} calculations (solid lines). Theoretical results underestimate the b+c process for $\epsilon_L \geq \SI{1.58}{eV}$, and overestimate it for $\epsilon_L=\SI{1.17}{eV}$.}
  \label{fig:2Dpeak_GWcalc}
\end{figure*}

To gain information whether the modification of relative intensity of the sub-peaks could be linked also to the strength of the EPC as a function of the exchanged momentum between the electron/hole and the phonon (fixed at each excitation energy by the resonance condition, see Fig.~\ref{fig:2Dpeak_exp}b) or only to the resonance condition met at the bottom of the split bands, we resort to a comparison with theoretical calculations. It is worth pointing out that the simple double-resonance Raman picture which neglects the dependence of the matrix elements on the electron and phonon momenta~\cite{PhysRevLett.85.5214} is not enough to reproduce the relative intensities of the sub-peaks. Indeed we have performed \emph{ab initio} calculations using the fourth-order Fermi golden rule following Ref.~\cite{venezuela2011theory}. In Figure~\ref{fig:2Dpeak_GWcalc} we compare the experimental 2D peak with the result of \emph{ab initio} calculations, performed following the methodology of Refs.~\cite{herziger2014bilayer, graziotto2023raman} and using the \texttt{EPIq} software~\cite{marini2023epiq}, at $\epsilon_L = \SI{2.33}{}, \SI{1.96}{}, \SI{1.58}{}$, and \SI{1.17}{eV}. We consider only these excitation energies since for higher $\epsilon_L$ the contribution of processes other than the a, b, c, d discussed above becomes more relevant, giving rise to an increased spectral weight at larger Stokes shift (as evident for the curve measured with $\epsilon_L = \SI{3.06}{eV}$), which also hinders the interpretation of the experimental sub-features. The calculations contain DFT only ingredients, apart from the rescaling of the electronic dispersion~\cite{venezuela2011theory} and the correction to the phonon bands~\cite{herziger2014bilayer} to match GW-corrected results. The inverse lifetime of the intermediate states is evaluated as in Ref.~\cite{herziger2014bilayer}. 

For each laser energy, we have calculated the Raman spectra at several different $\Delta\epsilon_\mathrm{split}$, which are tuned in \emph{ab initio} calculations by the appropriate modification of the carbon inter-layer distance (see Figs.~S5-6 in the SM). The total 2D peak and its single contributions a, b, c, d are shown in Fig.~\ref{fig:2Dpeak_GWcalc} for the value $\Delta\epsilon_\mathrm{split}=\SI{0.75}{eV}$. From this comparison one can see that these four double-resonance processes are indeed the dominant ones. Moreover, the good agreement in shape between the theoretical curves and the experimental ones for $\epsilon_L =$ \SI{2.33}{}, \SI{1.96}{}, \SI{1.58}{eV} supports the straightforward identification~\cite{herziger2014bilayer} of the experimental peak sub-features in terms of the processes depicted in Fig.~\ref{fig:schemeProcesses}a, b. Most notably, for these three spectra, the relative spectral weight of the different sub-features compares reasonably well between the experimental and theoretical curves. 
 By comparing the evolution of the theoretical spectral weight at several $\Delta\epsilon_\mathrm{split}$ with that extracted by fitting the experimental data (see Fig.~S6 in SM, and Fig.~\ref{fig:2Dpeak_GWcalc}i for $\Delta\epsilon_\mathrm{split} = \SI{0.75}{eV}$), we find that theory overestimates the spectral weight of the b+c process for $\epsilon_L$ = \SI{2.33}{}, \SI{1.96}{}, \SI{1.58}{eV} but always underestimates it at \SI{1.17}{eV}.
At $\epsilon_L = \SI{1.17}{eV}$ (Fig.~\ref{fig:2Dpeak_GWcalc}d, h) the theoretical line-shape fails in reproducing correctly the spectral weights of the two main features, i.e.\ the experimental peak is broader and more intense than the theoretical one. The extent of the underestimation of the b+c peak depends on the chosen $\Delta\epsilon_\mathrm{split}$. However, since the c process involves phonons having wavevectors closer to \textbf{K} with respect to the d ones (see Fig.~\ref{fig:2Dpeak_exp}b), one might wonder whether there is an underestimation of peak intensity that arises from a wrong evaluation of EPC for wavevectors closer to the \textbf{K} point within the DFT framework, in analogy to what found for monolayer graphene in Ref.~\cite{venanzi2023probing}. Therein it was argued that such underestimation could be due to the neglect, proper of DFT, of the dressing of the EPC by Coulomb interaction, but further theoretical investigation is needed to support the hypothesis in the present case, as it would also imply that the massiveness of the Dirac fermions involved in the resonant scattering processes is not a decisive limiting factor for the EPC enhancement.

%Indeed we find that if we multiply the contribution of process c by a factor of $2.8$ we recover an excellent agreement with the experimental spectrum (dashed black line in Fig.~\ref{fig:2Dpeak_GWcalc}f), in particular concerning the relative areas of the sub-features (an agreement between the peak heights would require a multiplication by a factor of $2.2$ instead). Being the area of the 2D peak processes dependent on the fourth power of the EPC~\cite{basko2008theory, venezuela2011theory} evaluated at the wavevector of the involved phonons, we find an increase of the EPC of a factor of $1.3$ when going from $\abs{\vb{q} - \mathbf{K}} = 0.21$ to \SI{0.14}{\angstrom^{-1}}. We remind that the EPC appears squared when dealing with electron-phonon scattering in transport problems, indeed here there are two phonons involved, from which the fourth power follows.

We can examine other possible causes of the theoretical underestimation of process c with respect to d. Although the magnitude of the contribution of processes a and b is dependent on $\Delta\epsilon_\mathrm{split}$, we can rule out the possibility that interference with process b can alone explain the enhanced intensity of the b+c sub-peak since at this energy also process a, which scales similarly to b, is extremely weak and indistinguishable from the D+D'' peak. Finally, calculations performed assuming a lower inverse lifetime of the intermediate states give the same results discussed above. \\

%\section{Conclusions}
We have shown that by means of resonance Raman spectroscopy with excitation in the infrared it is possible to suppress the contribution to the Raman spectrum of two of the most relevant scattering pathways building up the double-resonance 2D peak of bilayer graphene in the case of visible excitation. By comparing our experimental results to \emph{ab initio} calculations we identified the two most important processes contributing to the resonance 2D peak. We found that the relative intensity of the different scattering processes depends on how close to the band edge the electron-hole pair is excited. However, while experimentally the intensity of the b+c peak monotonically increases with respect to the d peak, DFT calculations always predict a decrease of the intensity of the b+c peak at $\epsilon_L = \SI{1.17}{eV}$. We suggest that such theoretical underestimation of the b+c intensity could stem from an enhancement of the electron-phonon coupling for the processes occurring closer to the \textbf{K} point. As a result, we believe that these findings demonstrate how infrared resonance Raman spectroscopy can provide more than a handful of insights into the physics of low-dimensional systems, and we remark that further experimental and theoretical studies are needed to elucidate the dependency of the EPC on the phonon wavevector and the role of Coulomb interactions. This would be of great relevance in order to model the impact of the electron-phonon coupling on the transport and optical properties of low-dimensional systems~\cite{novoselov20162d}.

%As a result, we believe that these findings demonstrate how infrared resonance Raman spectroscopy can provide more than a handful of insights into the physics of low-dimensional systems~\cite{novoselov20162d}, and could be used to investigate more exotic systems such as twisted bilayer graphene~\cite{jorio2013raman}, or transition-metal dichalcogenides~\cite{saito2016raman, sotgiu2022mose}.

\paragraph{Supporting Information}
Available online. Contains details on sample processing, laser sources, the definition of the \emph{Baskovian} fitting function, additional data taken at \SI{1.17}{eV} excitation energy on hBN-encapsulated bilayer graphene with back-gate voltage applied and on bilayer graphene on \ce{CaF2} substrate, details on the \emph{ab initio} calculations, the calculated Raman intensity resolved on the phonon wavevectors, calculations performed at different values of the energy split between bands 1-4, and the \emph{ab initio} calculated band structure of Bernal-stacked trilayer graphene.


\begin{acknowledgement}
We acknowledge financial support from PNRR MUR project PE0000023-NQSTI. We acknowledge the European Union's Horizon 2020 research and innovation program under grant agreements no.\ 881603-Graphene Core3 and the MORE-TEM ERC-SYN project, grant agreement no.\ 951215. We acknowledge PRACE for awarding us access to Joliot-Curie Rome at TGCC, France. Part of the calculations were performed on the DECI resource \emph{Mahti CSC} based in Finland at
\url{https://research.csc.fi/-/mahti}, and part on the ETH \emph{Euler} cluster at \url{https://scicomp.ethz.ch/}. L.G.\ acknowledges funding from the Swiss National Science Foundation (SNF project no.\ 200020-207795). T.O., B.B., and C.S.\ acknowledge funding from the Deutsche Forschungsgemeinschaft (DFG, German Research Foundation) under Germany’s Excellence Strategy—Cluster of Excellence Matter and Light for Quantum Computing (ML4Q) EXC 2004/1-390534769. Co-funded by the European Union (ERC, DELIGHT, 101052708). Views and opinions expressed are however those of the authors only and do not necessarily reflect those of the European Union or the European Research Council. Neither the European Union nor the granting authority can be held responsible for them.
\end{acknowledgement}

\bibliography{main}

\clearpage
\newpage

\begin{figure*}
    \begin{center}
        \includegraphics[width=\linewidth]{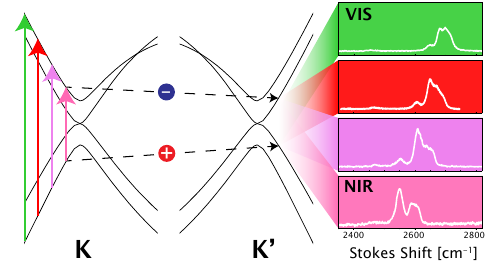}
    \end{center}
\end{figure*}

\end{document}

\maketitle
\noindent
\textbf{Sample description ---} We have investigated \ce{hBN}-encapsulated mono-, bi-, and trilayer graphene. The graphene is exfoliated from a graphite crystal which was purchased from \textit{HQ Graphene}. The \ce{hBN} flakes are exfoliated from crystals which were grown via a modified atmospheric pressure high temperature (APHT) process from an Fe flux as described in Refs.~\cite{EDGAR2014, ouaj2023hBN}. Boron (PI-KEM, 99.4-7\%) and iron (chemPUR, \SIrange{0.5}{1.5}{mm}, 99.97\%) were mixed and transferred to a rectangular alumina crucible
($5\times2\times2$ \SI{}{cm}) with lid and loaded into a tube furnace. The furnace was evacuated and refilled with \ce{N2} and \ce{H2} with qualities of 99.9999\% (6.0) for three times. The system was heated to \SI{1550}{\celsius} under a continuous flow of \ce{N2} (\SI{125}{sccm}) and \ce{H2} (5\% in \ce{Ar}, \SI{25}{sccm}) while maintaining a constant pressure of \SI{1.1}{bar}. After 24 hours the furnace was cooled to \SI{1400}{\celsius} at a rate of \SI{2}{\celsius/h}, to room temperature at \SI{300}{\celsius/h}, and subsequently heated to \SI{1550}{\celsius} again. After a dwelling time of 24 h, the system was cooled to \SI{1525}{\celsius} at a rate of \SI{0.5}{\celsius/h}, to \SI{1400}{\celsius} at a rate of \SI{2}{\celsius/h}, and lastly to room temperature at a rate of \SI{300}{\celsius/h}. The exfoliated flakes are dry-stacked on top of each other and the final stack is placed on a Cr/Au (\SI{5}{nm}/\SI{70}{nm}) pad and the heterostructure with the polycarbonate (PC) stamp is released from the polydimethylsiloxan (PDMS) at \SI{170}{\celsius} and subsequently placed in chloroform to remove the PC. The \ce{hBN}/graphene/\ce{hBN} stack is placed onto a gold pad to avoid photoluminescence from the silicon substrate at \SI{1.17}{eV} excitation energy, and which can also be employed to apply a gate voltage. The bottom and top \ce{hBN} layer are about \SI{45}{} and \SI{35}{nm} respectively. 

\noindent
\textbf{Laser sources ---} We employed an \ce{Nd}:YAG laser for the measurement at $\epsilon_L = \SI{1.17}{eV}$, \ce{HeNe} laser for the one at \SI{1.96}{eV}, solid-state lasers for the measurements at \SI{1.58}{}, \SI{2.33}{}, and \SI{3.06}{eV}, and an \ce{Ar} laser for the one at \SI{2.54}{eV}. 

\begin{figure}[hbt!]
    \centering
    \includegraphics[width=0.5\linewidth]{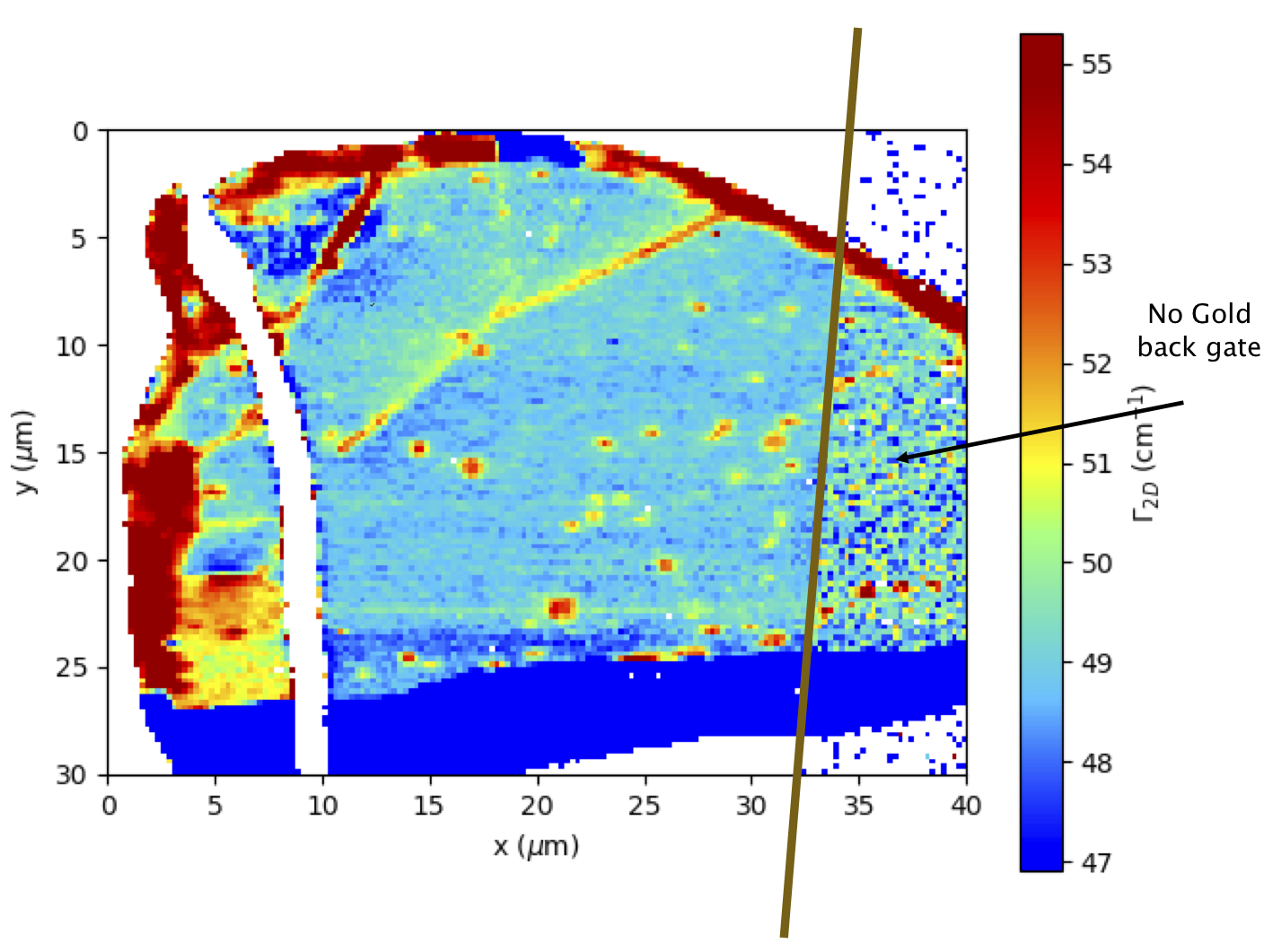}
    \caption{Raman map at $\epsilon_L = \SI{2.33}{eV}$ of the hBN-encapsulated bilayer graphene sample. The region on the left of the brown line is on top of the gold electrode. The color represents the width of the whole 2D peak (no fitting procedure has been employed). The spots in which the width increases indicate the presence of ripples or other kinds of defects.}
    \label{fig:BLRamanMap}
\end{figure}

\noindent
\textbf{Raman map of the bilayer sample ---} In Figure~\ref{fig:BLRamanMap} we report a Raman map acquired with \SI{2.33}{eV} excitation energy, showing the total full-width at half maximum (FWHM) of the 2D peak, estimated from the raw data without employing a fitting procedure. Measurements discussed in the main text were taken avoiding the regions with broadened 2D peak. 

\noindent
\textbf{Fitting function ---} The \emph{Baskovian} function $f_B(\omega)$, introduced in Ref.~\cite{basko2008theory}, is defined as follows 
\begin{equation}\label{eq:baskovian}
    f_B(\omega) = \frac{A_B \, \Gamma_B^2}{8(2^{2/3} - 1)} \frac{1}{\left( (\omega - \omega_0)^2 + \frac{\Gamma_B^2}{4(2^{2/3}-1)} \right)^{3/2}},
\end{equation}
where $\Gamma_B$ is the full-width at half maximum (FWHM) of $f_B$, $\omega_0$ is its central frequency, and $A_B$ is the total integrated area underneath it. The Baskovian function is more suited in describing the shape of the resonant features than the commonly used Lorentzian function~\cite{malard2007probing, cong2011raman}, but nonetheless they differ only in the decay of their tails. 

\begin{figure}
\centering
\includegraphics[width=1.\linewidth]{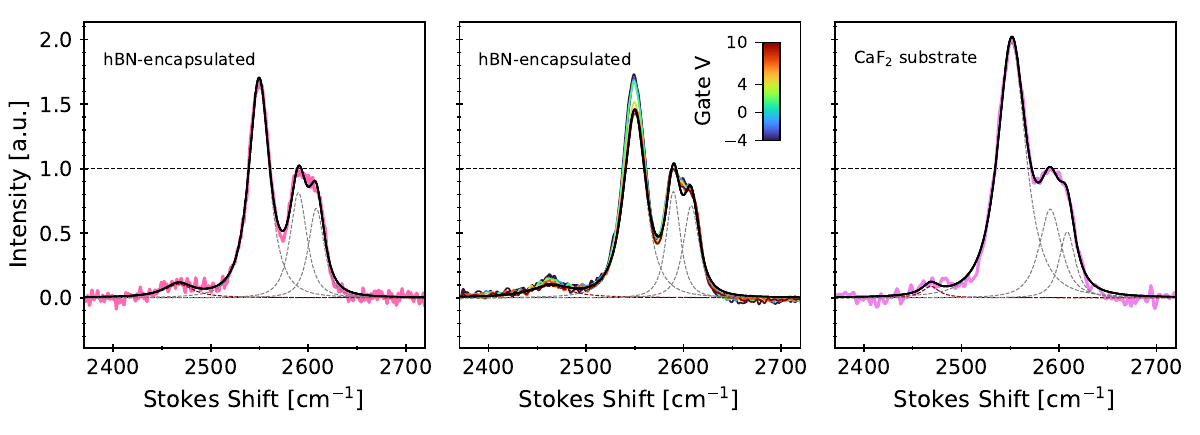}
\caption{Experimental spectra (solid coloured lines) at $\epsilon_L = \SI{1.17}{eV}$ for graphene samples in hBN encapsulation, in hBN encapsulation with gate voltage applied, and on \ce{CaF2} substrate. All the spectra have been normalized to the height of the d sub-peak. Observe how the ratio of the b+c to the d sub-peak varies when changing substrate/encapsulation. On the other hand, applying a gate voltage we do not observe substantial variations between \SI{-4}{} and \SI{4}{V}, and a decrease of the b+c peak intensity is observed only for voltages between \SI{4}{} and \SI{10}{V}. Notice also how the low-Stokes shift feature (a sub-peak or D+D'' double-resonance peak) changes. The black lines indicate the fitting function, which consists of the sum of one Lorentzian (to describe the a sub-peak or D+D'' peak) and three Baskovian functions (for the b+c, and d inner and outer sub-peaks), each represented with dashed lines. For the gated sample we report only the exemplary fitting at \SI{10}{V} back-gate.}
\label{fig:exp_spectra}
\end{figure}

\noindent
\textbf{Experimental dependence of the line-shape on the environment  ---}
We have studied the ratio of the b+c sub-peak to the d sub-peak (taken without further distinguishing inner or outer processes), with excitation energy $\epsilon_L = \SI{1.17}{eV}$, in samples within a different dielectric environment and by applying a backgate. Along with the hBN-encapsulated sample shown in the main text, we have investigated bilayer graphene on \ce{CaF2} substrate, and hBN-encapsulated bilayer graphene in the condition in which a back-gate voltage between \SI{-4}{} and \SI{10}{V} is applied. We report the experimental spectra in Figure~\ref{fig:exp_spectra}, where they have been normalized to the height of the d sub-peak. % which is less sensitive to the environment than the b+c sub-peak, since it involves the 2-3 bands, which without any external electric field applied are not split at the \textbf{K} point. 
We observe a sensible variation of the ratio of the b+c sub-peak to the d sub-peak across different samples and for gate voltages  above \SI{4}{V} (no variation is observed between \SI{-4}{} and \SI{4}{V}), along with a variation of the low-Stokes shift feature which we attribute to either the a sub-peak or to the D+D'' double-resonance peak, or to a mixture of the two. At 1.17 eV the intensity ratio of the b+c and d peaks appears to be thus extremely sensitive to doping, strain or dielectric environment, but we can not discriminate among these causes as probably the effect on the peak intensity stems from a combination of several physical factors. 

\begin{figure}
\centering
\includegraphics[width=0.6\linewidth]{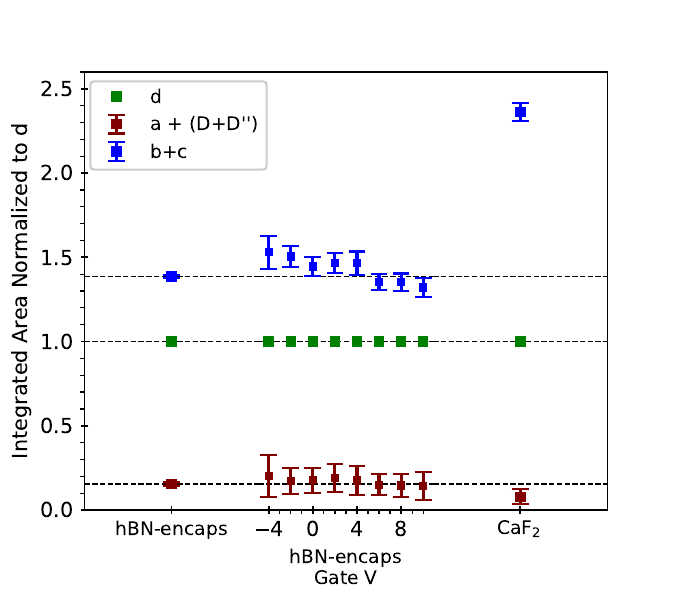}
\caption{Integrated areas of the Lorentzian/Baskovian functions employed in the fitting procedure, normalized to the combined area of the two Baskovian function which describe the d sub-peak (inner and outer), for different bilayer samples. The dashed horizontal lines are located at the values obtained for the ungated hBN-encapsulated sample, which is discussed in the main text.}
\label{fig:exp_spectra_fitresults}
\end{figure}

In order to characterize the variation of the ratio of the sub-peaks to the d sub-peak, we perform a fitting of the spectra via a curve given by the sum of a Lorentzian function (which describes the a sub-peak or D+D'' peak) and three Baskovian functions (which describe the b+c sub-peak, and the d sub-peak with the further distinction in inner and outer processes, respectively). We report the results of the fitting procedure as dashed lines in Fig.~\ref{fig:exp_spectra}. In Figure~\ref{fig:exp_spectra_fitresults} we report the integrated areas of the aforementioned Lorentzian/Baskovian functions normalized to the integrated area of the two Baskovian functions which describe the d sub-peak (inner and outer), for the samples described above.

\noindent
\textbf{Details on the \emph{ab initio} calculations ---} The Raman scattering intensity reads~\cite{venezuela2011theory}
\begin{equation}\label{eq:RamanIntensity}
    I(\omega) = \frac{1}{N_{\vb{q}}} \sum_{\vb{q},\nu,\mu} I_{\nu\mu} (\vb{q}) \,\delta\left(\omega_L - \omega- \omega_{-\vb{q}}^\nu - \omega_{\vb{q}}^\mu\right),
\end{equation}
where $\omega_{\vb{q}}^\mu$ is the frequency of a phonon with wavevector $\vb{q}$ belonging to mode $\mu$, and $I_{\nu\mu} (\vb{q}) = \abs{\sum_{\vb{k}, \mathrm{process}} K_\mathrm{process}(\vb{k},\vb{q},\nu,\mu)}^2/N_{\vb{k}}^2$ is the probability of exciting two phonons with opposite wavevectors $\vb{q}$, given in terms of the matrix elements of the different a, b, c, d processes $K_\mathrm{process}$, which depend upon the electron-phonon coupling, the electron-light interaction, the electron dispersion, and only slightly on the phonon dispersion~\cite{venezuela2011theory, graziotto2023raman}, such that one can substitute it with a constant (i.e.\ the TO phonon frequency at the \textbf{K} point) without affecting the value of $K_\mathrm{process}$. The unscreened electric dipole and the screened electron-phonon matrix elements are calculated from first principles~\cite{giannozzi2009quantum} in LDA approximation~\cite{PhysRevB.23.5048} on a $36\times36$ electron momentum grid and on a $6\times6$ phonon momentum grid~\cite{PhysRevB.82.165111, mostofi2014updated}, and then Wannier-interpolated~\cite{herziger2014bilayer, marini2023epiq} on a telescopic non-uniform electron momentum grid densified around the resonant region, and a uniform phonon momentum grid around the \textbf{K} point. The electronic and phonon dispersion are calculated from first principles and GW corrected as in Ref.~\cite{herziger2014bilayer}. The Raman spectrum of an individual process $I_\mathrm{process}(\omega)$ is obtained theoretically by dropping the summation inside the square modulus (i.e.\ by considering only one $K_\mathrm{process}$) in Eq.~\ref{eq:RamanIntensity}. Its integrated area is given by
\begin{equation}
    A_\mathrm{process} = \int_\mathrm{2D} I_\mathrm{process}(\omega) d\omega,
\end{equation}
and one can see that the delta function in Eq.~\ref{eq:RamanIntensity} can be integrated out, so that $A_\mathrm{process}$ does not depend on $\omega_{-\vb{q}}^\nu, \omega_{\vb{q}}^\mu$, i.e.\ on the phonon dispersion. This result stays true also when substituting the delta function with a Lorentzian, which keeps into account the finite lifetime of the final phonon states, since the latter is still normalized to one when the full interval of the 2D peak is taken into account. 

\begin{figure}[hbt!]
\includegraphics[width=\textwidth]{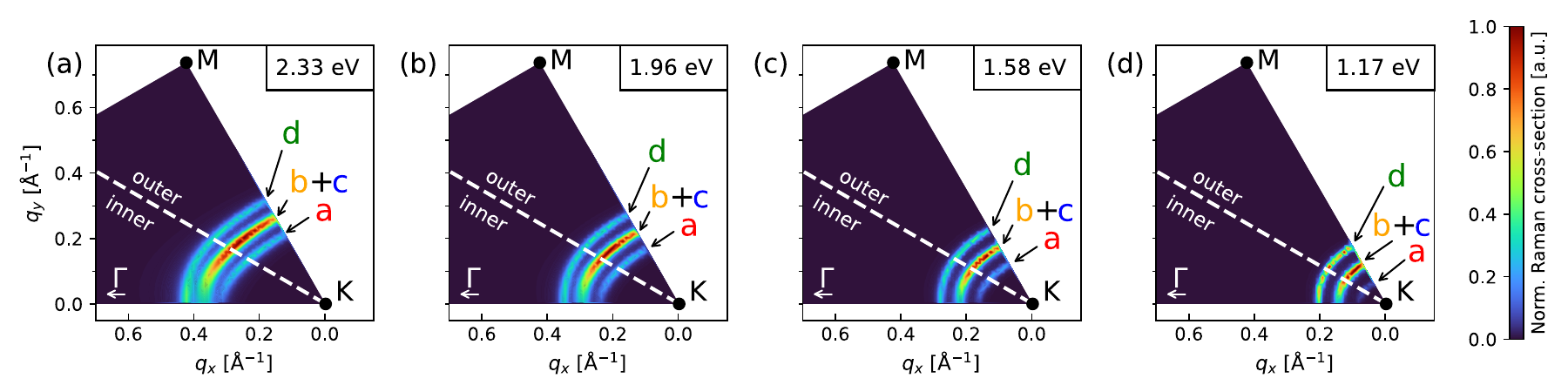}
\caption{Raman intensity as a function of phonon wavevector $\vb{q}$ in the irreducible wedge of the FBZ, for $\epsilon_L = \SI{2.33}{}, \SI{1.96}{}, \SI{1.58}{}, \SI{1.17}{eV}$ (left to right) and $\Delta\epsilon_\mathrm{split}=\SI{0.75}{eV}$. The colored letters indicate the a, b, c, d processes which different phonon wavevectors contribute to. We further indicate the attribution of inner (below the dashed line) and outer (above) contributions, following Ref.~\cite{herziger2014bilayer}, which extends the simplified labelling along the $\boldsymbol{\Gamma}-\mathbf{K}-\mathbf{M}$ to the whole BZ.}
\label{fig:Iq_FBZ}
\end{figure}

\noindent
\textbf{Intensity as a function of phonon wavevector on the whole BZ ---} In Figure~\ref{fig:Iq_FBZ} we display the 2D peak Raman intensity as a function of phonon wavevector, reduced to the irreducible wedge of the FBZ, for $\epsilon_L = \SI{2.33}{}, \SI{1.96}{}, \SI{1.58}{}$, and \SI{1.17}{eV}, that is $\mathcal{I}(\vb{q}) = \sum'_{\nu,\mu} I_{\nu\mu}(\vb{q})$ where the summation is restricted to the energy window of the 2D peak. We identify the contributions to processes a, b, c, d stemming from different phonon wavevectors, and indicate the extension of the inner/outer labelling to the whole BZ, following Ref.~\cite{herziger2014bilayer}.

\begin{figure}[hbt!]
    \centering
    \includegraphics[width=\textwidth]{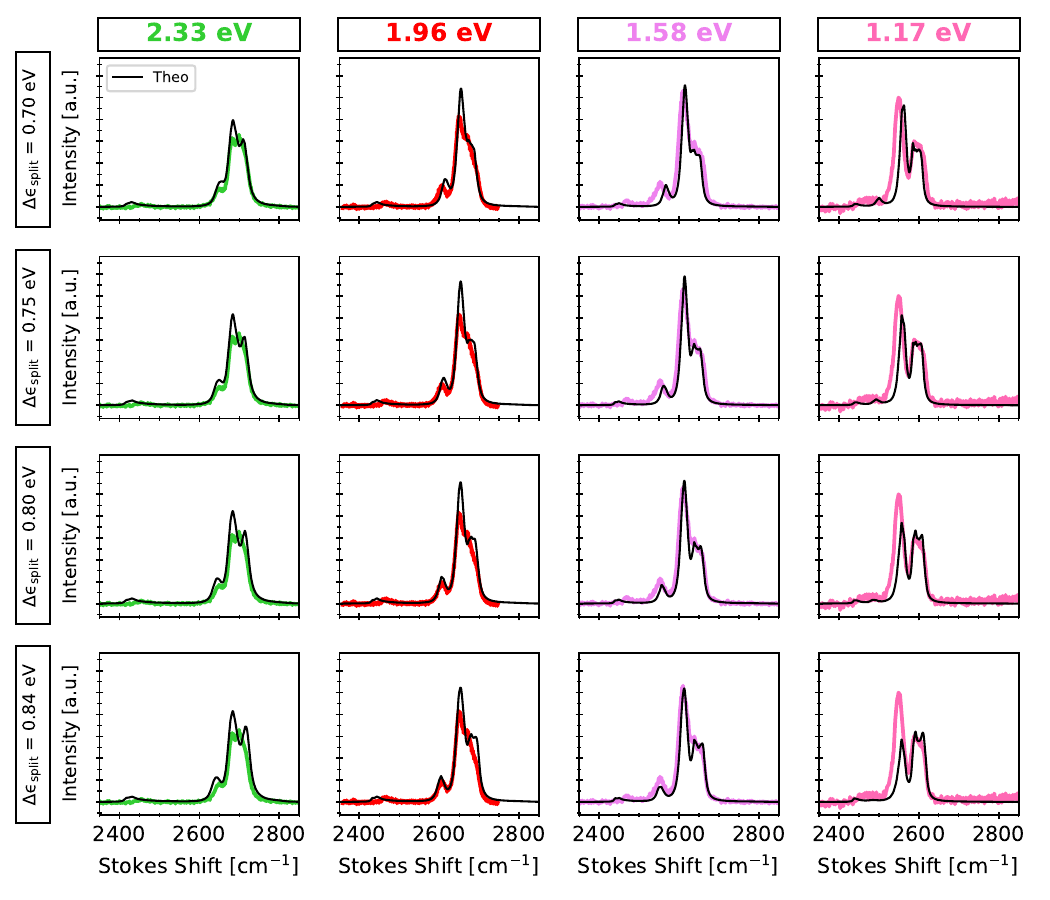}
    \caption{Comparison between theoretical calculations (black solid lines) performed by varying $\Delta\epsilon_\mathrm{split}$ (the energy gap between the split 1-4 bands of bilayer graphene) and experimental spectra (colored solid lines), for different excitation energies. Each spectrum is normalized to the integrated area of its d sub-peak. The excitation energy is varied along the rows, while $\Delta\epsilon_\mathrm{split}$ is varied, for the theoretical spectra only, along the columns. In each column (i.e.\ for the same excitation energy) all the panels report the same experimental spectrum. As mentioned in the main text, the theoretical curves have been rigidly shifted in order to overlap with the experimental spectra.}
    \label{fig:allspectra_allgaps}
\end{figure}

\noindent
\textbf{Calculations with different $\Delta\epsilon_\mathrm{split}$ ---} In order to investigate the impact on the 2D line-shape of the energy gap $\Delta\epsilon_\mathrm{split}$ between the split 1-4 bands of bilayer graphene, we have repeated the ab initio calculation of the Raman spectrum varying the distance between the graphene planes, which corresponds to changing $\Delta\epsilon_\mathrm{split}$ from \SI{0.70}{} to \SI{0.84}{eV}, values which are consistent with the ones obtained experimentally~\cite{kuzmenko2009infrared, PhysRevB.78.235408}. Notice that the variation of the interlayer distance needed to obtain a \SI{50}{meV} change in the splitting is of order 1\%, below experimental precision detection. We compare the different theoretical calculations with the experimental spectra in Figure~\ref{fig:allspectra_allgaps}.

\begin{figure}
    \centering
    \includegraphics[width=0.9\linewidth]{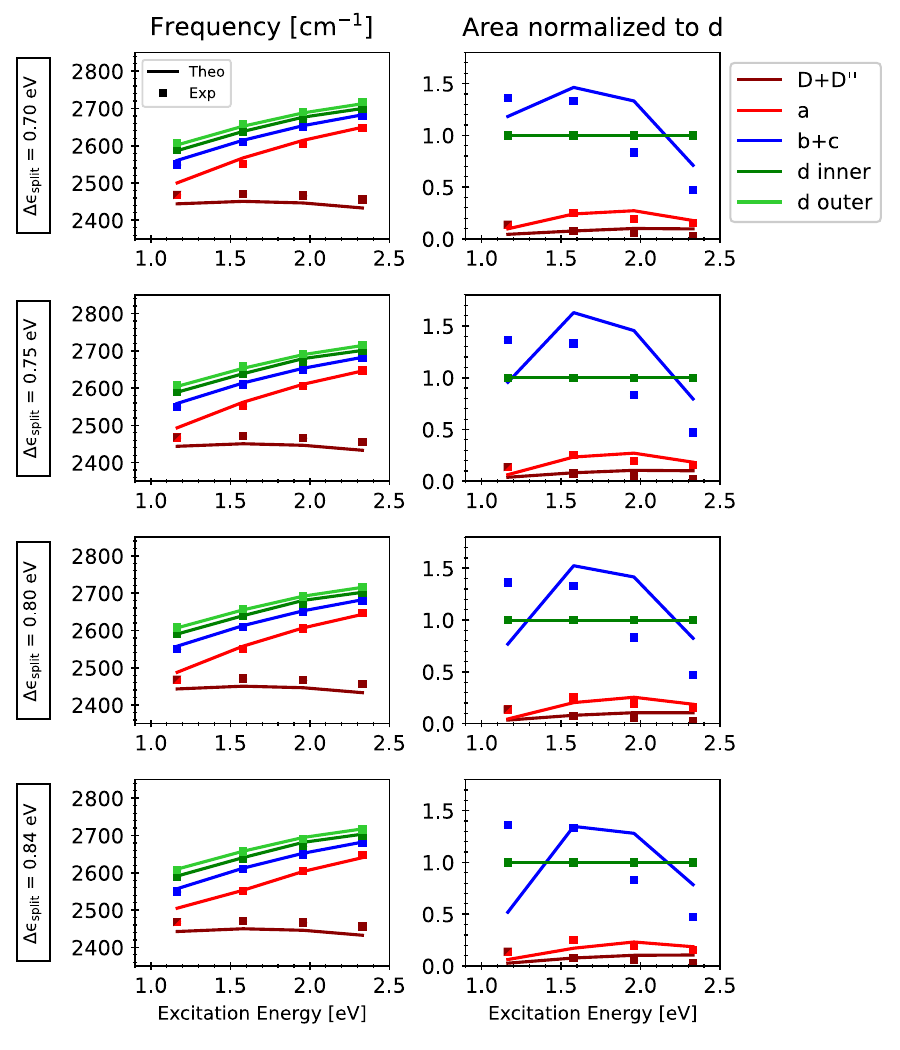}
    \caption{Parameters obtained from the fitting procedure of both the theoretical (solid lines) and experimental (squares) spectra reported in Figure~\ref{fig:allspectra_allgaps}. $\Delta\epsilon_\mathrm{split}$ is varied along the columns for the theoretical spectra only. The first column reports the central frequency of the a, b, c, d sub-peaks (color-coded following the legend on the right, with the further distinction in inner and outer for the d process) and of the double resonance D+D'' peak as a function of the excitation energy. The second column reports the integrated area of the a, b, c, d sub-peaks and of the double resonance D+D'' normalized to the total area of the d sub-peak (considering both inner and outer processes).}
    \label{fig:fits_diffgapsandlasers}
\end{figure}

By performing a fitting with a curve given by the sum of a Lorentzian function (which describe the D+D'' peak) and four Baskovian functions (which describe the a sub-peak, the b+c sub-peak, and the d sub-peak with the further distinction in inner and outer processes, respectively) on both the experimental and theoretical spectra, we can directly compare the relative intensities of the different processes. We report the results of the fitting procedure in Figure~\ref{fig:fits_diffgapsandlasers}. Noteworthy, the experimental intensity of the b+c monotonically grows by lowering $\epsilon_L$ while the theoretical behavior predicts a decrease of the b+c intensity at 1.17 eV with respect to that at 1.58 eV. Such theoretical underestimation of the b+c intensity could be explained in terms of a joint effect of bandgap renormalization and enhancement of electron phonon coupling for electron hole pairs closer to $\mathbf{K}$.

\begin{figure}[hbt!]
    \centering
    \includegraphics[width=0.5\linewidth]{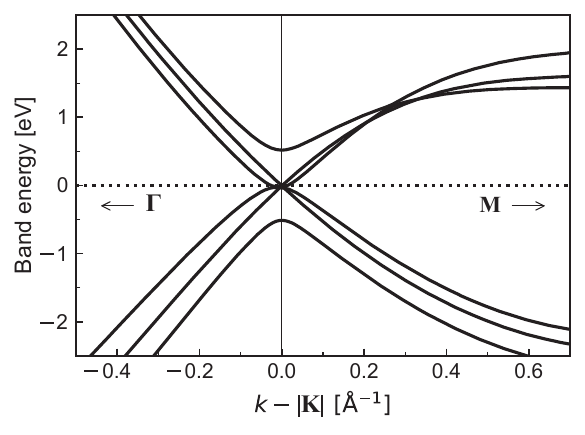}
    \caption{DFT ab initio calculated bands of Bernal-stacked (ABA) trilayer graphene, as a function of the electron wavevector $\vb{k}$ along the $\boldsymbol{\Gamma}-\mathbf{K}-\mathbf{M}$ direction.}
    \label{fig:trilayer-bands}
\end{figure}

\noindent
\textbf{Trilayer graphene band structure ---} In Figure~\ref{fig:trilayer-bands} we report the electronic band dispersion of Bernal-stacked (ABA) of trilayer graphene, as obtained from DFT ab initio calculations.

\clearpage
\bibliography{main}